\documentclass[12pt]{article}
\pdfoutput=1
\usepackage{amsmath}
\usepackage[dvips]{graphicx}
\usepackage[T1]{fontenc}
\DeclareGraphicsExtensions{.png,.pdf}

\usepackage{color}
\input amssym

\def\red#1{{\color{black} #1}}

\begin{document}

\def\prg#1{\par\medskip\noindent{\bf #1}}  \def\ra{\rightarrow}
\newcounter{nbr}
\def\note#1{\bitem\vspace{-5pt}\addtocounter{nbr}{1}
            \item{} #1\vspace{-5pt}
            \eitem}
\def\lra{\leftrightarrow}              \def\Ra{\Rightarrow}
\def\nin{\noindent}                    \def\pd{\partial}
\def\dis{\displaystyle}                \def\Lra{{\Leftrightarrow}}
\def\grl{{GR$_\Lambda$}}               \def\vsm{\vspace{-8pt}}
\def\cs{{\scriptstyle\rm CS}}          \def\ads3{{\rm AdS$_3$}}
\def\Leff{\hbox{$\mit\L_{\hspace{.6pt}\rm eff}\,$}}
\def\bull{\raise.25ex\hbox{\vrule height.8ex width.8ex}}
\def\ric{{Ric}}                        \def\tmgl{\hbox{TMG$_\Lambda$}}
\def\Lie{{\cal L}\hspace{-.7em}\raise.25ex\hbox{--}\hspace{.2em}}
\def\sS{\hspace{2pt}S\hspace{-0.83em}\diagup}
\def\hd{{^\star}}                      \def\dis{\displaystyle}
\def\mb#1{\hbox{{\boldmath $#1$}}}     \def\kn#1{\hbox{KN$#1$}}
\def\ul#1{\underline{#1}}              \def\phb{\phantom{\Big|}}
\def\nb{~\marginpar{\bf\Large ?}}      \def\ph{\phantom{xxx}}
\def\bb{\bar b}                        \def\bom{\bar\om}
\def\bH{\bar H}
\def\ir#1{{}^{(#1)}}
 \def\tgr{{GR$_\parallel$}}
\def\hook{\hbox{\vrule height0pt width4pt depth0.3pt
\vrule height7pt width0.3pt depth0.3pt
\vrule height0pt width2pt depth0pt}\hspace{0.8pt}}
\def\inn{\hook}
\def\first{\rm (1ST)}  \def\second{\hspace{-1cm}\rm (2ND)}
\def\ppl{{pp${}_\Lambda$}}

\def\G{\Gamma}        \def\S{\Sigma}        \def\L{{\mit\Lambda}}
\def\D{\Delta}        \def\Th{\Theta}       \def\Ups{\Upsilon}
\def\a{\alpha}        \def\b{\beta}         \def\g{\gamma}
\def\d{\delta}        \def\m{\mu}           \def\n{\nu}
\def\th{\theta}       \def\k{\kappa}        \def\l{\lambda}
\def\vphi{\varphi}    \def\ve{\varepsilon}  \def\p{\pi}
\def\r{\rho}          \def\Om{\Omega}       \def\om{\omega}
\def\s{\sigma}        \def\t{\tau}          \def\eps{\epsilon}
\def\nab{\nabla}      \def\btz{{\rm BTZ}}   \def\heps{{\hat\eps}}
 \def\vth{\vartheta}

\def\bR{\bar{R}}      \def\bT{\bar{T}}     \def\hT{\hat{T}}
\def\tG{{\tilde G}}   \def\cF{{\cal F}}    \def\cA{{\cal A}}
\def\cL{{\cal L}}     \def\cM{{\cal M }}   \def\cE{{\cal E}}
\def\cH{{\cal H}}     \def\hcH{\hat{\cH}}  \def\cT{{\cal T}}
\def\hA{\hat{A}}      \def\hB{\hat{B}}     \def\hK{\hat{K}}
\def\cK{{\cal K}}     \def\hcK{\hat{\cK}}  \def\cT{{\cal T}}
\def\cO{{\cal O}}     \def\hcO{\hat{\cal O}} \def\cV{{\cal V}}
\def\tom{{\tilde\omega}}  \def\cE{{\cal E}} \def\bH{\bar{H}}
\def\cR{{\cal R}}    \def\hR{{\hat R}{}}   \def\hL{{\hat\L}}
\def\tb{{\tilde b}}  \def\tA{{\tilde A}}   \def\hom{{\hat\om}}
\def\tT{{\tilde T}}  \def\tR{{\tilde R}}   \def\tcL{{\tilde\cL}}
\def\he{{\hat e}}    \def\hom{{\hat\om}}   \def\hth{\hat\theta}
\def\hxi{\hat\xi}    \def\hg{\hat g}       \def\hb{{\hat b}}
\def\tH{{\tilde H}}  \def\tV{{\tilde V}}   \def\bA{\bar{A}}
\def\bV{\bar{V}}     \def\bxi{\bar{\xi}}
\def\knl{\text{KN}$(\l)$}   \def\mknl{\text{KN}\mb{(\l)}}
\def\bPhi{\bar\Phi}
\def\chm{\checkmark}                \def\chmr{\red{}}
\def\tt{\tilde t}
\vfuzz=2pt 
\def\nn{\nonumber}
\def\be{\begin{equation}}             \def\ee{\end{equation}}
\def\ba#1{\begin{array}{#1}}          \def\ea{\end{array}}
\def\bea{\begin{eqnarray} }           \def\eea{\end{eqnarray} }
\def\beann{\begin{eqnarray*} }        \def\eeann{\end{eqnarray*} }
\def\beal{\begin{eqalign}}            \def\eeal{\end{eqalign}}
\def\lab#1{\label{eq:#1}}             \def\eq#1{(\ref{eq:#1})}
\def\bsubeq{\begin{subequations}}     \def\esubeq{\end{subequations}}
\def\bitem{\begin{itemize}}           \def\eitem{\end{itemize}}
\renewcommand{\theequation}{\thesection.\arabic{equation}}
\title{Extremal Kerr black hole entropy in Poincar\'e gauge theory}

\author{B. Cvetkovi\'c and D. Rakonjac\footnote{
   Email addresses: \texttt{cbranislav@ipb.ac.rs, danilo.rakonjac@ipb.ac.rs}} \\
Institute of Physics, University of Belgrade,\\
                     Pregrevica 118, 11080 Belgrade-Zemun, Serbia}
\date{}
\maketitle

\begin{abstract}
We analyze the near horizon symmetry of the extremal Kerr black hole within the framework of Poincar\'e gauge theory (PG) for two
important limiting cases: Riemannian and teleparallel solution. We show that the algebra of canonical generators is realized by
Virasoro algebra, with central charge which depends on the black hole horizon radius . The conformal entropy of the black hole is obtained via Cardy formula.
\end{abstract}
\section{Introduction}
Recently a new Hamiltonian method  \cite{bc-2019} for the computation of black hole entropy within the framework
of Riemann-Cartan geometry has been proposed. The method has been verified for a number of \red{vacuum} solutions
such as \red{Schwarzschild(-AdS), Kerr(-AdS)} solution as well as \red{a solution coupled to electromagnetic field}, Kerr-Newmann-AdS solution, \cite{bc-2019,bc-2019a,bc-2020b,bc-2022a,bc-2022b}.

The method \cite{bc-2019} is based on a variational principle, originally proposed by Regge and Teitelboim, see \cite{regge}.
The  black hole entropy is obtained from the variation of the boundary term on the black hole horizon, i.e.
$T\d S=\d \G_H$. In the framework of Riemaninnan geometry this method was established and developed by Wald
\cite{wald}.
 Moreover, the differentiability
of the canonical generator is closely related to the  validity of the first law of the black hole mechanics.

\red{The method \cite{bc-2019} is  inapplicable  in the case of the extremal black holes.} Namely,  in that case black hole
temperature vanishes\red{, $T=0$,} and the equation $T\d S=\d\G_H$ cannot be solved for the black hole entropy.
For extremal black holes the first law is satisfied disregarding the value of the black hole entropy.
However in general relativity (GR), there is another way of computing black hole entropy of the extremal Kerr black holes
based on near horizon conformal symmetry, \red{in regard of the recently established Kerr/CFT correspondence} \cite{guica}, see also \cite{carlip}.

The subject of the present paper is the computation of the black hole entropy for the extremal Kerr black holes in the framework of PG,
where both curvature and torsion influence the gravitational dynamics \cite{hehl,mb-1,mb-2},
\red{by analyzing the near horizon conformal symmetry.} Let us note
that near horizon structure of black holes with torsion has already been examined within threedimensional gravity \cite{simic}.

After introducing the suitable \red{set} of \red{consistent} near horizon boundary conditions for extremal Kerr black holes in  PG,
\red{we obtain that asymptotic symmetry  group has a conformal
subgroup, realized by Virasoro algebra.}
We shall show that the first order formulation of the generator of the local symmetry derived in \cite{bc-2019}, can be used to compute the near horizon algebra of the improved generators, as well as the corresponding central charge. \red{These results
are used to compute conformal entropy of the extremal Kerr black hole via Cardy's formula. The result for the entropy represents a smooth limit of the result for the gravitational black hole entropy of the generic (non-extremal) Kerr black hole.} Thus, we  demonstrated the full power of the Nester's covariant Hamiltonian approach \cite{nester}. \red{ and   contributed to the better understanding
of the equality of gravitational and conformal entropy.}

The paper is organized as follows. In section 2 we shall introduce the tetrad formulation of the extremal Kerr black hole solution
\red{and near horizon geometry (NHEK) in the framework of PG. The suitable set of consistent asymptotic conditions for near horizon geometry
in tetrad formalism of PG is established in section 3. Inspection of the symmetries that
preserve these boundary conditions leads to conformal symmetry of NHEK geometry.}
In the section 4 we shall consider the canonical realization of the near horizon conformal symmetry for Riemannian
solution in PG. \red{We shall make use of the  canonical generator from the first order formulation obtained in \cite{bc-2019} to compute the conserved and  central charge of the conformal near horizon symmetry, which depends on black hole horizon radius. The conserved and central charge are going to be used in the Cardy's formula  to compute the conformal black hole entropy.}
Another important limiting case of PG, teleparallel gravity, \red{where gravitational dynamics is characterized by vanishing curvature and non-vanishing torsion,} is analyzed in section  5.
Section 6 is devoted to concluding remarks, while appendices contain some technical details.

Our conventions are the same as in ref. \cite{bc-2022b}. The Latin indices $(i,j,\dots)$ are the local Lorentz indices, the Greek indices $(\m,\n,\dots)$ are the coordinate indices, and both run over $0,1,2,3$. The orthonormal coframe (tetrad) $\vth^i$ and the metric compatible (Lorentz) connection $\om^{ij}=-\om^{ji}$  are 1-forms, the dual basis (frame) is $e_i=e_i{}^\m\pd_\m$.
The metric components in the local Lorentz and coordinate basis are $\eta_{ij}=(1,-1,-1,-1)$ and $g_{\m\n}=\eta_{ij} \vth^i{}_\m\vth^j{}_\n$, respectively, and $\ve_{ijmn}$ is the totally antisymmetric symbol with $\ve_{0123}=1$. The Hodge dual of a form $\alpha$ is denoted by $\hd\alpha$, and the wedge product of forms is implicit.

\section{Tetrad formulation of extremal  Kerr black hole geometry}
\setcounter{equation}{0}

In this section we shall introduce the tetrad formulation of the extremal Kerr black hole geometry.
\red{We shall introduce the near horizon geometry (NHEK),} which represents our starting point in the study of the near horizon structure of the extremal Kerr black holes within Riemann-Cartan geometry.
\pagebreak
\red{\subsection{Metric, conserved charges and the first law}}
\red{Let us now give a brief overview of the basic features of the extremal Kerr black holes. We use the same
notation as in \cite{bc-2019a}.}
A ``diagonal" form of the extremal Kerr metric ($m=a$) in Boyer-Linquist coordinates \cite{griffiths}
\bsubeq
\be
ds^2=N^2\Big(dt+m\sin^2\th d\phi\Big)^2-\frac{dr^2}{N^2}
  -\r^2d\th^2-\frac{\sin^2\th}{\r^2}\Big[m dt
                                       +(r^2+m^2)d\phi\Big]^2\,,
\ee
where
\bea
N=\frac{r-m}{\r}\, \qquad \r^2:=r^2+m^2\cos^2\th\, .
\eea
\esubeq
The \red{equation $N=0$ defines the extremal black hole horizon, which radius $r_+$ is given by}
\be
r_+=m\,.
\ee
\red{The two horizons that exist in Kerr solution coincide in the extremal case.}
The value of the angular velocity $\Om_+$ on the horizon takes the simple form
\be
\Om_+=\frac{1}{2r_+}\equiv \frac1{2m}\, ,
\ee
while the surface gravity and temperature vanish
\be
\k=\frac{r_+-m}{2mr_+}=0\,,\quad T=\frac{\k}{2\pi}=0\,.
\ee
The conserved charges  energy and angular momentum of Kerr black hole
in PG take the form \cite{bc-2019a}
\bea
E=m\,,\qquad J=m^2\,.
\eea
Let us note that the first law of black hole thermodynamics reads

\be
0=T\d S\equiv \d E-\Om_+\d J=\d m-\frac1{2m}\d m^2\,.
\ee

It is satisfied in the of the extremal Kerr black hole {\it disregarding the value of the black hole entropy}.

\red{\subsection{Near horizon extremal Kerr geometry}} Let us now introduce the following coordinate
transformations \cite{guica, carlip}
\bea
\tilde t=\frac{\ve t}{2r_+}\,,\qquad y=\frac{\ve r_+}{r-r_+}\,,\qquad \vphi=\phi+\Om_+t\,,
\eea
\red{along with the limit  $\ve\ra 0$.}  The metric takes the following form
\be\lab{2.8}
d s^2=r_+^2(1+\cos^2\th)\left[\frac {d\tilde t^2}{y^2}-\frac{dy^2}{y^2}-d\th^2-\left(\frac{2\sin\th}{1+\cos^2\th}\right)^2\left(d\vphi-\frac{d\tilde t}y\right)^2\right]\,.
\ee
\red{Let us  note that the above transformation is not a simple coordinate transformation, in
that the resulting geometry is not equivalent to the starting one. That can be readily
seen by noticing that the above metric is not asymptotically flat. The resulting geometry
has been extensively studied in \cite{guica,bardeen}}
\prg{Tetrads.} The form of the metric \lab{2.8} implies the  following "diagonal" choice of the vielbein $\vth^i$
\bea\lab{2.9}
&&\vth^0=\frac{\r_+}{y}d\tilde t\,,\qquad \vth^1=-\frac{\r_+}y dy\,,\nn\\
&&\vth^2=\r_+d\th\, ,\qquad
 \vth^3=\frac{2\sin\th}{\sqrt{1+\cos^2\th}}r_+\left(d\vphi-\frac{d\tilde t}{y}\right)\, ,
\eea
where $\r_+=r_+\sqrt{1+\cos^2\th}$

For later convenience let us now also display the form of the Riemannian connection and curvature.
\prg{Riemannian connection.} From the relation $d \vth^i+\tom^i{}_k \vth^k=0$, we get that the
nonzero components of the Riemannian connection are
\bea
&&\tom^{01}=-\frac{\vth^0}\r_+-\frac{r_+^2\sin\th}{\r_+^3} \vth^3\,,\quad \tom^{02}=\frac{r_+^2\sin2\th}{2\r_+^3}\vth^0\,,\qquad
\tom^{12}=\frac{r_+^2\sin2\th}{2\r_+^3}\vth^1\,,\nn\\
&&\tom^{03}=-\frac{r_+^2\sin\th }{\r_+^3}\vth^1\,,\qquad
\tom^{13}=-\frac{r_+^2\sin\th }{\r_+^3}\vth^0\,,\qquad \tom^{23}=\frac{2r_+^2\cot\th}{\r_+^3}\vth^3\,.
\eea
\prg{Riemannian curvature.} The Riemannian curvature reads
\bea
&&\tR^{01}=-2C \vth^0\vth^1-2D\vth^2\vth^3\,,\qquad \tR^{02}=C\vth^0\vth^2-D\vth^1\vth^3\,,\nn\\
&& \tR^{03}=C\vth^0\vth^3+D\vth^1\vth^2\,,\qquad\tR^{12}=C\vth^1\vth^2-D\vth^0\vth^3\,,\nn\\
&&\tR^{13}=C\vth^1\vth^3+D\vth^0\vth^2\,,\qquad \tR^{23}=-2C\vth^2\vth^3+2D\vth^0\vth^1\,,
\eea
where
$$
C=\frac{r_+^4}{\rho_+^6}(1-3\cos^2\th)\,,\quad D=\frac{r_+^4\cos\th}{\r_+^6}(3-\cos^2\th)\,.
$$
\section{Asymptotic conditions and asymptotic symmetry}
\setcounter{equation}{0}

In this section we shall first introduce the  asymptotic conditions for the metric of the extremal Kerr black hole solution in the near horizon region. \red{Due to the fact that NHEK is
not asymptotically flat, finding consistent asymptotic boundary conditions is not a priori an
obvious task. This result has been established in GR \cite{guica}, however, care should be taken,
as asymptotic boundary conditions of the metric do not precisely dictate the asymptotic
boundary conditions for the tetrad. Therefore the consistent choice for the tetrad, as well
as near horizon conformal symmetry will be provided in this section.}
\prg{Metric asymptotics.} In accordance with \cite{guica} we introduce the following set of consistent asymptotic conditions for the metric near the asymptotic boundary $y=0$
\bea
g_{\m\n}\sim\left(\ba{cccc}
 \cO_{-2}&\cO_0&\cO_1&\bar g_{\tt\vphi}+\cO_0\\
\cO_0&\bar g_{yy}+\cO_{-1}&\cO_0&\cO_{-1}\\
\cO_1&\cO_0&\bar g_{\th\th}+\cO_1&\cO_1\\
\bar g_{\tt\vphi}+\cO_0&\cO_{-1}&\cO_1&\cO_0
\ea\right)\,,
\eea
where
\bea
&&\bar g_{\tt\tt}=\frac{r_+^2(1+\cos^2\th)}{y^2} - \frac{4\sin^2\th}{1+\cos^2\th}\frac {r_+^2}{y^2}\,,\nn\\
&&\bar g_{\tt\vphi}=\frac{4\sin^2\th}{1+\cos^2\th}\frac {r_+^2}y\,,\nn\\
&&\bar g_{yy}=-\frac{r_+^2(1+\cos^2\th)}{y^2}\,,\nn\\
&&\bar g_{\th\th}=-r_+^2(1+\cos^2\th)\,,\nn\\
\eea
are background metric components and we use the notation $\cO_n:=\cO(y^n)$.
\prg{Tetrad fields.} The asymptotic form of the vielbein is given by
\bea
\vth^i{_\m}\sim \left(
\ba{cccc}
\cO_{-1}&\cO_1&\cO_2&\cO_1\\
\cO_1&\bar \vth^1{_y}+\cO_0&\cO_1&\cO_0\\
\cO_1&\cO_0&\bar \vth^2{_\th}+\cO_1&\cO_1\\
\bar \vth^3{_t}f(\vphi)+\cO_0&\cO_1&\cO_2&\displaystyle\frac{\bar \vth^3{_\vphi}}{f(\vphi)}+\cO_1
\ea
\right)\,,
\eea
where background tetrad fields are given by
\bea
\bar \vth^i{_\m}=\left(
\ba{cccc}
\dis\frac{\r_+}y&0&0&0\\
0&-\dis\frac{\r_+}y&0&0\\
0&0&\r_+&0\\
-\dis\frac{2\sin\th r_+}{\sqrt{1+\cos^2\th}y}&0&0&\dis\frac{2\sin\th r_+}{\sqrt{1+\cos^2\th}}
\ea
\right)\,,
\eea
where $f(\vphi)=1+h(\vphi)$  is an arbitrary function of  $\vphi$, such that $h(\vphi)\ll 1$.
\prg{Asymptotic symmetry.}
The transformation law of  $\vth^i{_\mu}$ under PG transformations reads
$$
\d_0 \vth^i{_\mu}=\th^i{_k}\vth^k{_\m}-(\pd_\mu \xi^\r)\vth^i{_\r}-\xi^\r\pd_\r \vth^i{_\m}\,,
$$
where $\xi^\mu$ and $\th^{ij}$ are parameters of local translations and local Lorentz rotations, respectively.

The asymptotic form of the metric \red{is preserved by asymptotic Killing vector  $\xi^\m$ of the
following form}
\bsubeq\lab{3.5}
\bea
\xi^{\tilde{t}}=T+\cO_3\,,\qquad\xi^y=y\pd_\vphi\eps(\vphi)+\cO_2\,,\qquad \xi^\theta = \cO_1\, , \qquad \xi^\vphi=\eps(\vphi)+\cO_2\,.\lab{3.5a}
\eea
\red{The transformation corresponding to $T$ is a constant time translation, and we are able to
restrict our attention to the conformal group disregarding this transformation, due to the
fact that its generator commutes with the generator of the conformal symmetry. The subdominant terms correspond to trivial diffeomorphisms, and  they be disregarded, so that the final form of the asymptotic Killing vector reads
\be
\xi=\left(y\pd_\vphi\eps(\vphi)\right)\pd_y+\eps(\vphi)\pd_\vphi\,.
\ee}

All the parameters of Lorentz rotations obtained from the invariance of the tetrad fields are asymptotically vanishing
\bea
&&\th^{01}=\cO_2\,,\quad\th^{02}=\cO_2\,,\quad \th^{03}=\cO_1\,,\nn\\
&&\th^{12}=\cO_1\,,\qquad \th^{13}=\cO_2\,,\quad \th^{23}=\cO_2\,. \lab{3.5b}
\eea
\esubeq

The Riemannian connection can be expressed in terms of tetrad fields and therefore its asymptotic form is
invariant under transformations \eq{3.5}.

The transformations with $\eps=0$ represent {\it residual gauge transformations}
which give trivial contribution to the conserved charge. Therefore, the  asymptotic symmetry group is defined as a factor
group with respect to residual transformations. From the general algebra of PG we get the composition rule for the
asymptotic transformations
\bea\lab{3.7}
&&[\d_0(\eps_1),\d_0(\eps_2)]=\d_0(\eps_3)\,,\nn\\
&&\eps_3=\eps_1\eps'_2-\eps_2\eps'_1\,,
\eea
where $\eps':=\pd_\vphi\eps$.
In terms of Fourirer modes
$$
\ell_n:=\d_0(\eps=e^{in\vphi})\,,
$$
the algebra of the asymptotic symmetry takes the  Virasoro form
\be
[\ell_n,\ell_m]=i(m-n)\ell_{m+n}\,.
\ee
In what follows we shall analyze the canonical realization of the asymptotic symmetry in the
two important cases -- Riemannian PG solution and teleparallel solution.

\section{Riemannian extremal Kerr black hole in PG}
\setcounter{equation}{0}

In this section we shall analyze the Riemannian extremal Kerr black hole within the  framework of PG. \red{It is well known that Kerr black hole is a solution of the GR field equations with vanishing cosmological constant $\L = 0$, and so is its extremal case. Existence of the
corresponding near horizon geometry is a property of the extremal Kerr, and it should be
noted that this near horizon geometry is the solution of GR field equations as well \cite{kunduri}.} From the general theorem which states that GR solutions also represent solutions of PG, one can conclude that the Kerr black hole, \red{as well as NHEK} also satisfy,  the PG field equations for $\L=0$.  There is also a direct proof based on the form of the effective PG Lagrangian \cite{bc-2019a}
\be
L_G=-\hd(a_0 R+2\L)+\frac{1}{2}b_1 R^{ij} \hd R_{ij}\,,           \lab{4.1}
\ee
which defines the corresponding covariant momenta as $H_i=0$ and
\bsubeq
\be
H_{ij}=-2a_0\hd(\vth_i\vth_j)+b_1\hd R_{ij}\, ,
\ee
or in more detail
\bea
&&H_{01}=-2a_0\vth^2\vth^3+2b_1(-2C\vth^2\vth^3+2D\vth^0\vth^1)\, ,                   \nn\\
&&H_{02}=~~2a_0\vth^1\vth^3+2b_1(-C\vth^1\vth^3-D\vth^0\vth^2)\, ,                    \nn\\
&&H_{03}=-2a_0\vth^1\vth^2+2b_1(C\vth^1\vth^2-D\vth^0\vth^3)\, ,                    \nn\\
&&H_{12}=-2a_0\vth^0\vth^3+2b_1(C\vth^0\vth^3+D\vth^1\vth^2)\,,                      \nn\\
&&H_{13}=~~2a_0\vth^0\vth^2+2b_1(-C\vth^0\vth^2+D\vth^1\vth^3)\,,                    \nn\\
&&H_{23}=-2a_0\vth^0\vth^1+2b_1(-2C\vth^0\vth^1-2D\vth^2\vth^3)\,.
\eea
\esubeq
\red{ In the following subsections, the conserved and central charge on the horizon will be
computed.}
We shall make use of the general expression for the variation of the canonical generator on the horizon \cite{bc-2019}
\bsubeq\lab{4.3}
\bea
&&\d\G_H=\oint_{S_H} \d B(\xi)\,,                                  \\
&&\d B(\xi):=(\xi\inn \vth^{i})\d H_i+\d \vth^i(\xi\inn H_i)
   +\frac{1}{2}(\xi\inn\om^{ij})\d H_{ij}
   +\frac{1}{2}\d\om^{ij}(\xi\inn\d H_{ij})\, .
\eea
\esubeq
where $\xi$ is either exact or an asymptotic Killing vector.
\subsection{Conserved charge}
 Let us now compute the conserved charge on the horizon. It is obtained for $\xi=\pd_\vphi$. Since $H_i=0$  the variation of the canonical
generator \eq{4.3} reduces now to:
\bea
&& \d\G_H:=\oint_{S_H}\d B\,,\nn\\
&&\d B=\frac 12\om_{ij\vphi}\d H^{ij}+\frac12(\d \om^{ij})H_{ij\vphi}\,.
\eea
The non-vanishing contribution to the conserved charge
stems from:
\be
\tom^{01}{_\vphi}\d H_{01}=\frac{4a_0\sin^2\th}{(1+\cos^2\th)^2}\d(2\sin\th r_+^2)d\th d\vphi\,.
\ee
Now we get that the conserved charge reads:
\bea
J&=&\oint_{S_H}\tom^{01}{_\vphi}\d H_{01}=16\pi a_0 r_+^2\equiv r_+^2\,,
\eea
where we used
$$
\int_0^\pi\frac{\sin^3\th}{(1+\cos^2\th)^2}d\th=1\,.
$$

\subsection{Central charge and black hole entropy}

We shall compute the central charge from the algebra of the improved canonical generators, which has the following
form:
\be\lab{4.7}
\{\tG(\eps_1),\tG(\eps_2)\}=\tG(\eps_3)+C\,,
\ee
where $\eps_3$ is defined by the composition rule \eq{3.7} and  $C$ is  the central term of the algebra.

After using the main result of the seminal Brown-Henneaux paper \cite{brown}, the canonical algebra \eq{4.7} can be simplified and it takes the form of  the following weak equality:
\bea
\{\tG(\eps_1),\tG(\eps_2)\}\approx \d_0(\eps_1)\G_H(\eps_2)\approx \G_H(\eps_3)+C\,,
\eea
The central term is a constant functional and therefore it can be computed by varying the background configuration.
\red{Non-zero contributions to the above variation are given by
\bea
	\frac12\oint_{S_H} (\xi_2 \inn \bom^{ij})\d_0(\xi_1)\bH_{ij} + \d_0(\xi_1)\bom^{ij}(\xi_2\inn \bH_{ij})&=& 8a_0r^2_+\int_0^{2\pi}(\eps_1\eps'_2-\eps_2\eps'_1)d\vphi \nn\\
&-& 4a_0r^2_+\int_0^{2\pi}(\eps_1'\eps_2''-\eps_2'\eps_1'')d\vphi\,
\eea
The first term in the equation above can be identified with the surface term with parameter $\eps_3 = \eps_1\eps'_2-\eps_2\eps'_1$, while the second one gives the central charge}
\be
C=-4a_0r_+^2\int_0^{2\pi}(\eps'_1\eps''_2-\eps'_2\eps''_1)d\vphi\,.
\ee
For the computational details see appendix A.

In terms of Fourier modes the canonical algebra of the improved generators reads:
\be
\{L_n,L_m\}=-i(n-m)L_{m+n}-\frac c{12}in^3\d_{n,-m}\,,
\ee
where in the string theory normalization
\be
c=12\cdot 16\pi a_0 r_+^2=12r_+^2\equiv 12 J\,.
\ee
Let us note that central charge does not depend on action paramater $b_1$, but does depend on the
parameter of the horizon radius $r_+$.

Now the entropy can be calculated via Cardy's formula:
\be
S=2\pi\sqrt{\frac{c}6\left(J-\frac{c}{24}\right)}=2\pi r_+^2\,.
\ee
The result for the conformal entropy of the extremal Riemannian Kerr black hole in PG represents
a smooth limit obtained from the expression for gravitational entropy of the generic Kerr black hole
in the same theory \cite{bc-2019a}.

\section{Extremal Kerr black hole in TG}
\setcounter{equation}{0}
Teleparallel gravity is a special case of PG, which is defined by the condition of vanishing Riemann-Cartan curvature, $R^{ij}=0$ \cite{nitch}.
The Kerr solution does indeed solve the equations of motion of teleparallel gravity \cite{bc-2019a}.

 Let us note that from  the condition  $R^{ij}=0$ it does not follow that the   connection $\om^{ij}$  (which is a "pure gauge") vanishes.
 Since, connection does not influence the PG dynamics, we shall adopt the simplest choice $\om^{ij}=0$. Thus, the tetrad field remains the only dynamical variable, and torsion takes the form $T^i=d \vth^i$. For the spacetime with tetrad \eq{2.9}, the nonvanishing components of  torsion are given by
\bea
&&T^0=-\frac1{\r_+}\vth^0\vth^1+\frac{r_+^2\sin 2\th}{2\r_+^3}\vth^0\vth^2\,,
\qquad T^1=\frac{r_+^2\sin 2\th}{2\r_+^3}\vth^1\vth^2\,,\nn\\
&&T^3=\frac{2r_+^2\sin\th}{\r_+^3}\vth^0\vth^1
       +\frac{2r_+^2\cos\th}{\r_+^3\sin\th}\vth^2\vth^3\,.
\eea
All three irreducible parts of $T^i$ are nonvanishing.

The Lagrangian of the teleparallel equivalent of GR, so called \tgr\  is given by
\be
L_T:=a_0T^i\,\hd\left(\ir{1}T_i-2\ir{2}T_i-\frac 12\ir{3}T_i\right)\,.
\ee
 This equivalence ensures that every vacuum solution of GR is also a solution of \tgr\ and in particular, this is true for the extremal Kerr spacetime. Though the two theories are  dynamically equivalent, their geometric content  is quite different: GR is characterized by a Riemannian curvature and vanishing torsion, whereas the teleparallel geometry of \tgr\ has a nontrivial torsion but vanishing curvature.

The covariant momentum is given by
\bsubeq
\be
H^i=2a_0\,\hd\Big(\ir{1}T^i-2\ir{2}T^i-\frac{1}{2}\ir{3}T^i\Big)\,,
\ee
and its explicit form reads
\bea
H^0&=&2a_0\left(-\frac{r^3_+\sin\th}{\r_+^3}\vth^0\vth^2+\frac{\cos\th}{\r_+\sin\th}\vth^1\vth^3\right)\,,                                \nn\\
H^1&=& 2a_0\left(\frac{\cos\th}{\r_+\sin\th}\vth^0\vth^3
             -\frac{r_+^2\sin\th}{\r_+^3}\vth^1\vth^2\right)\,,                   \nn\\
H^2&=&-2a_0\frac{\vth^0\vth^3}{\r_+}\,,                 \nn\\
H^3&=&2a_0\left(\frac{r_+^2\sin 2\th}{\r_+^3}\vth^0\vth^1+\frac1\r_+\vth^0\vth^2-\frac{r_+^2\sin\th}{\r_+^3}\vth^2\vth^3\right)\,.\quad
\eea
\esubeq

\subsection{Conserved charge}
 We shall  now compute the conserved charge on the horizon. It is obtained for $\xi=\pd_\vphi$. Since $H_{ij}=0$  the variation of the canonical
generator \eq{4.3} reduces now to
\bea
&& \d\G_H:=\oint_{S_H}\d B\,,\nn\\
&&\d B=b_{i\vphi}\d H^i+(\d \vth^i) H_{i\vphi}\,.
\eea
The non-vanishing contribution to the conserved charge
stems from
\be
\vth^3{_\vphi}\d H_3+(\d \vth^3) H_{3\vphi}=\frac{4a_0\sin^2\th}{(1+\cos^2\th)^2}\d(2\sin\th r_+^2)d\th d\vphi\,.
\ee
Now we get that the conserved charge reads
\bea
J&=&\oint_{S_H}\vth^3{_\vphi}\d H_3+(\d \vth^3) H_{3\vphi}=16\pi a_0 r_+^2\equiv r_+^2\,.
\eea

\subsection{Central charge and black hole entropy}

The central charge can again be obtained from the  algebra of the improved canonical generators
as in the previous section.
The central term, computed by varying the background configuration,  (for details see appendix B) is given by
\be
C=-4a_0r_+^2\int_0^{2\pi}(\eps'_1\eps''_2-\eps'_2\eps''_1)d\vphi\,.
\ee

In terms of Fourier modes the canonical algebra of the improved generators reads:
\be
\{L_n,L_m\}=-i(n-m)L_{m+n}-\frac c{12}in^3\d_{n,-m}\,,
\ee
where in the string theory normalization
\be
c=12\cdot 16\pi a_0 r_+^2=12r_+^2\equiv 12 J\,.
\ee

The entropy of the extremal Kerr black hole in \tgr\ can be calculated via Cardy's formula:
\be
S=2\pi\sqrt{\frac{c}6\left(J-\frac{c}{24}\right)}=2\pi r_+^2\,.
\ee
The result for the conformal entropy of the extremal  Kerr black hole in \tgr\ represents
a smooth limit obtained from the expression for gravitational entropy of the generic Kerr black hole
in the same theory \cite{bc-2019a}.
\section{Concluding remarks}

We analyzed the near horizon symmetry for the extremal Kerr black hole in the framework of PG by using
the Hamiltonian approach in the first order formulation of the theory. \red{We
have shown, considering two important limits of PG, namely Riemannian and teleparallel
solution, that the algebra of improved canonical generators for the extremal Kerr black
hole takes the form of Virasoro algebra with classical central charge which depends on the
black hole horizon radius.}
We computed the extremal Kerr black hole entropy  via Cardy's formula, \red{finding that conformal entropy calculated this way equals the smooth limit of the
non-extremal gravitational entropy.}
The method we developed can be extended to the extremal Kerr black hole with torsion in the generic case. Also it would be interesting to
examine near horizon structure of the extremal Reissner-Nordstr\"om-like black hole solutions with torsion  \cite{jorge-2017,jorge-2018}.

\section*{Acknowledgments}

This work was  supported by the Ministry of Education, Science and Technological Development of the Republic of Serbia.

\appendix
\section{Central charge for extremal Riemannian black hole in PGT}
\setcounter{equation}{0}
The central charge stems from the variation of the
surface term $\d_0(\eps_1)\G_H(\eps_2)$ on the background configuration.

Asymptotic Killing vector, after disregarding residual gauge transformations reads
\be
\xi = y\eps'\pd_y + \eps\pd_\vphi\,.
\ee
We shall make use of the following non-vanishing internal products
\bsubeq
\bea
&&\xi\inn\vth^1 = -r_+\sqrt{1+\cos^2\th}\eps'\,,\qquad \xi\inn \bar \vth^3 = \frac{2r_+\sin\th}{\sqrt{1+\cos^2\th}}\eps\,,\\
&&\xi\inn\bar\om^{01} = -\frac{2\sin^2\th}{(1+\cos^2\th)^2}\eps\,,\qquad \xi\inn\bar\om^{03}=\frac{\sin\th}{1+\cos^2\th}\eps'\,,\nn\\
&&\xi\inn\bar\om^{12} = -\frac{\sin\th\cos\th}{1+\cos^2\th}\eps'\,,\qquad \xi\inn\bar\om^{23}=\frac{4\cos\th}{(1+\cos^2\th)^2}\eps\,.
\eea
\esubeq
The non-vanishing terms in the variation of the background configuration of tetrad fields (on the boundary defined
by $\tilde t=\rm const$ and $y\ra 0$) are given by
\bsubeq
\bea
&& \d_0\bar\vth^1 =r_+\sqrt{1+\cos^2\th}\eps''d\varphi\,,\\
&& \d_0\bar\vth^3 = -\frac{2r_+\sin\theta}{\sqrt{1+\cos^2\theta}}\eps'd\vphi\,.
\eea
\esubeq
Consequently, for the canonical  momenta $\bH_{ij}$ we obtain
\bsubeq
\bea
&&\d_0\bH_{01} =4(a_0 + 2b_1C)r^2_+\sin\theta\eps'd\th d\vphi\,, \\
&&\d_0\bH_{03} =2(a_0 -b_1C)r^2_+(1+\cos^2\th)\eps''d\th d\vphi\,,\\
&&\d_0\bH_{12} = -2b_1Dr^2_+(1+\cos^2\th)\eps''d\th d\vphi\,, \\
&&\d_0\bH_{23} =  8b_1Dr^2_+\sin\th\eps'd\th d\vphi\,.
\eea
\esubeq
After term by term integration we get
\bsubeq
\bea
&&\oint_{S_H} (\xi_2 \inn \bom^{01})\d_0(\xi_1)\bH_{01} = -\left(8a_0r^2_++\frac12 b_1(8+3\pi)\right)\int_0^{2\pi}\eps_2\eps_1'd\vphi\,, \\
&& \oint_{S_H} (\xi_2 \inn \bom^{03})\d_0(\xi_11)\bH_{03} = \left(4a_0r^2_+-b_1\right)\int_0^{2\pi}\eps_2'\eps''_1d\vphi\,,\\
&& \oint_{S_H}(\xi_2 \inn \bom^{12})\d_0(\xi_1)\bH_{12} = b_1\int_0^{2\pi}\eps_2'\eps''_1d\vphi\,, \\
&& \oint_{S_H} (\xi_2 \inn \bom^{23})\d_0(\xi_1)\bH_{23} = \frac{1}{2} b_1(8+3\pi)\int_0^{2\pi}\eps_2\eps'_1d\vphi\,.
\eea
\esubeq
In the above expression we made use of the following definite integrals
\bea
&& \int_0^\pi \frac{\sin^3\th}{(1+\cos^2\th)^2} = 1\,,\qquad \int_0^\pi \frac{\sin^2\th(1-3\cos^2\th)}{(1+\cos^2\th)^5} = \frac{1}{32}(8 + 3\pi),,\nn \\
&& \int_0^\pi \frac{\sin\th(1-3\cos^2\th)}{(1+\cos^2\th)^3} = \frac{1}{2}\,,\qquad \int_0^\pi\frac{\sin\th\cos^2\th(3-\cos^2\th)}{(1+\cos^2\th)^3} = \frac{1}{2}\,, \\
&& \int_0^\pi \frac{\sin\th\cos^2\th(3-\cos^2\th)}{(1+\cos^2\th)^5} = \frac{1}{64}(8 + 3\pi)\,.\nn
\eea

After summing up all the contributions we get the first term

\be
\frac12\oint_{S_H} (\xi_2 \inn \bom^{ij})\d_0(\xi_1)\bH_{ij}= -8a_0r^2_+\int_0^{2\pi}\eps_2\eps'_1d\vphi + 4a_0r^2_+\int_0^{2\pi}\eps_2'\eps_1''d\vphi\,.
\ee
The non-vanishing variations of the connection are given by
\bsubeq
\bea
&& \d_0\bom^{01} = \frac{2\sin^2\th}{(1+\cos^2\th)^2}\eps'd\vphi\,, \\
&& \d_0 \bom^{03} = -\frac{\sin\th}{1+\cos^2\th}\eps''d\vphi\,, \\
&& \d_0\bom^{12}  = \frac{\sin\th\cos\th}{1+\cos^2\th}\eps''d\vphi\,, \\
&& \d_0 \bom^{23} = -\frac{4\cos\th}{(1+\cos^2\th)^2}\eps'd\vphi\,.
\eea
\esubeq
The internal products with canonical momenta are given by
\bsubeq
\bea
&&\xi\inn\bH_{01} = 4(a_0+2b_1C)r_+^2\sin\th\eps d\th\,, \\
&&\xi\inn\bH_{03} = 2(a_0 - b_1C)(1+\cos^2\th)r_+^2\eps'd\th\,, \\
&&\xi\inn \bH_{12} = -2b_1D(1+\cos^2\th)r_+^2\eps'd\th,, \\
&&\xi\inn \bH_{23} = 8b_1Dr_+^2\sin\th\eps d\th\,.
\eea
\esubeq
Since all the integrals over $\theta$ are identical the second term takes the following form
\bea
\frac{1}{2}\d_0(\xi_1)\om^{ij}(\xi_2\inn H_{ij})=8a_0r^2_+\int_0^{2\pi}\eps_1\eps'_2d\vphi -4a_0r^2_+\int_0^{2\pi}\eps_1'\eps_2''d\vphi\,.
\eea

\section{Central charge of the extremal Kerr black hole in TG}
\setcounter{equation}{0}
In TG the curvature equals zero and therefore we have to compute the variation
the variation of the background canonical covariant momenta $\bH_i$
\bsubeq
\bea
&&\d_0 \bH_1 = -2a_0 \frac{r_+\sin\th}{\sqrt{1+\cos^2\th}}\eps''d\th d\vphi\,, \\
&&\d_0 \bH_3 = -4a_0\frac{\sin^2\th}{r_+(1+\cos^2\th)^{\frac 32}}\eps'd\th d\vphi\,.
\eea
\esubeq
As in the Riemannian case after performing integration $\th$, we directly obtain
\be
(\xi_2\inn\bar\vth^i)\d_0(\xi_1) \bH_i = 4a_0r^2_+\int_0^{2\pi}\eps_2'\eps_1''d\vphi - 8a_0r^2_+\int_0^{2\pi}\eps_2\eps_1'd\vphi\,.
\ee
The non-trivial internal products of the canonical momenta read
\bsubeq
\bea
&&\xi\inn\bH_1 = -2a_0 \frac{r_+\sin\th}{\sqrt{1+\cos^2\th}}\eps'd\th\,, \\
&&\xi\inn\bH_3 = -4a_0\frac{\sin^2\th}{r_+(1+\cos^2\th)^{\frac32}}\eps d\th\,.
\eea
\esubeq
The second term takes the following form
\be
\d_0(\xi_1)\vth^i(\xi_2\inn\bH_i)=-4a_0r^2_+\int_0^{2\pi}\eps_1'\eps_2''d\vphi + 8a_0r^2_+\int_0^{2\pi}\eps_1\eps_2'd\vphi\,.
\ee

\end{document}